# Optimizing the Nonlinear Optical Response of Plasmonic Metasurfaces

Yael Blechman, Euclides Almeida, Basudeb Sain, Yehiam Prior
*Department of Chemical and Biological Physics, Weizmann Institute of Science, Rehovot 76100, Israel*
*yael.blechman@weizmann.ac.il, yehiam.prior@weizmann.ac.il*

# Abstract

The nonlinear optical response of materials to exciting light is enhanced by resonances between the incident laser frequencies and the energy levels of the excited material. Traditionally, in molecular nonlinear spectroscopy one tunes the input laser frequencies to the molecular energy levels for highly enhanced doubly or triply resonant interactions. With metasurfaces the situation is different, and by proper design of the nanostructures, one may tune the material energy levels to match the incoming laser frequencies. Here we use multi-parameter genetic algorithm methodologies to optimize the nonlinear Four Wave Mixing response, and show that the intuitive conventional approach of trying to match the transmission spectrum to the relevant laser frequencies indeed leads to strong enhancement, but not necessarily to the optimal design. We demonstrate, experimentally and by direct nonlinear field calculations, that the near field mode distribution and spatial modes overlap are the dominant factor for optimized design.

# Introduction

From the early days of nonlinear optics, interactions such as Raman or wave mixing have been studied in gas-phase molecules or solid-state crystalline systems, in which the energy levels are fixed, and in order to achieve resonance enhancements one tunes the laser frequencies to the molecular resonances [1]. Advances in the design and fabrication of nano structures open up the possibility to do the reverse: namely for fixed laser frequencies, one tunes the 'system' for resonance enhancement. The resonant local field enhancements, induced by surface plasmons, greatly reduce the interaction length typically required for creating a strong nonlinear effect [2]. This significant scale down in the overall size of nonlinear devices enables their integration into a nano-photonic circuit, thus, leading the way for new nanoscale nonlinear optical components [3–5]. Nanoscale nonlinear devices can also be useful for applications such as sensing [6], super-resolution imaging [7,8], light manipulation [9–11], and in terahertz generation [12].

In arrays of metallic particles, once the separation between the particles exceeds 20-30 nm, the rapid decay of the evanescent field outside the metal causes the spectral response of the array to be essentially identical to that of a single particle. Metasurfaces consisting of arrays of nanometric cavities in thin metal films, however, are very different: the localized surface plasmons (LSP) generated at and near a cavity couple to propagating surface plasmons (SPPs) in the metal film, which, in turn, couple back to local fields in the neighboring cavity, giving rise to a strongly coupled set of individual sources. These nanohole arrays provide high versatility in designing their linear response, consisting of both localized (LSP), propagating (SPP) and cavity modes. Several works have investigated nonlinear enhancement in metallic nanohole arrays [13–21], however, not many studied third-order processes, and even fewer considered the question of optimizing these processes.

The dependence of the linear optical properties of nanohole arrays on hole size, shape and array parameters has been extensively studied[22-37]. The Extra Ordinary Transmission (EOT) through a nanocavity array, as shown by Ebbesen [22,23], when normalized to the cavities area, can even be larger than unity for certain structures and at well-defined frequencies. The high EOT at a given frequency results from plasmonic resonances (SPP or LSP), indicating high local field enhancement at this frequency. These local enhancements, if present at the frequencies of the nonlinear process, are equivalent to the resonance enhancement due to tuning the lasers to molecular energy levels. The naïve approach to optimization of nonlinear optical processes derives from this analogy, and one expects the nonlinear optical process, i.e. Four Wave Mixing (FWM), to be maximal when all three input fields are enhanced by properly designed nanocavity arrays.

Recently, based on the naïve approach described above, we demonstrated an order of magnitude enhancement of the FWM response of an array of rectangular cavities (figure 1) by tuning the response of the array to have linear transmission peaks at the pump and anti-Stokes input frequencies. In that experiment, for the sake of transparency of the process, of several possible parameters, only the aspect ratio of rectangular holes (maintained at a given area) was optimized[21]. Proper design of the metasurface led to a double resonance, showing transmission peaks at both input frequencies, but not at the generated FWM frequency. The question is whether an optimization process that takes into account all the available parameters of the metasurface will provide better optimization, and more generally, is the naïve approach emulating results from atomic and molecular spectroscopy indeed the best optimization strategy.

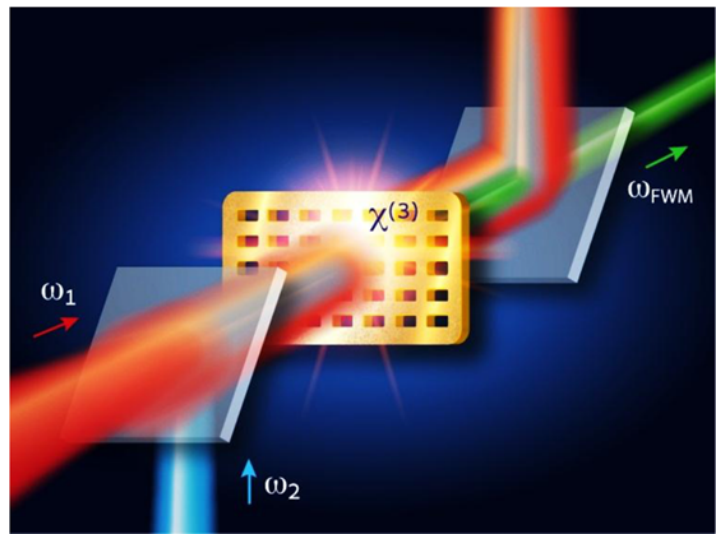

Figure 1 Illustration of the FWM process in a nanohole array (Reproduced from [21]).

In what follows, we discuss two approaches to the optimization of nonlinear FWM. The first is an extension of our earlier naïve approach, but with including all the geometrical parameters of the system – hole dimensions and array geometry, as well as the film properties. Here we will find the parameter set that gives rise to a triple resonance situation, where transmission peaks are found at all three relevant frequencies, input as well as generated. In parallel, we introduce a different approach, which is based on our ability to directly calculate the nonlinear FWM signal for a given geometry. Using FTDT calculations, we show that by using the Genetic Algorithm we are able to find a configuration that is non-intuitive, do not conform to the conventional wisdom of mandatory transmission peaks at the relevant frequencies. We discuss both approaches and compare the predictions to experiments. In this paper, we consider Four Wave Mixing from an array of nanocavities, but the same methodology is applicable to other NL processes and other geometries.

# Results

Consider the FWM process $\omega_{FWM} = 2\omega_1 - \omega_2$, where the main input frequency $\omega_1$ is chosen to match the Ti:Sapphire laser frequency at 800 nm and the goal is to maximize the FWM signal at 622nm. This choice of frequencies dictates the second input beam to be at 1127nm. The process is illustrated in figure 1. Our metasurface (MS) consists of rectangular arrays of nano-sized rectangular holes milled in a thin freestanding gold film (see Figure 2). There are two reasons why we opted for free standing gold films; The first and more important one is the need to eliminate the nonresonant FWM signal from the much thicker substrate (i.e. glass), which, due to the quadratic dependence on the propagation length would swamp the signal from the metasurface. The second reason is the better coupling of light to the MS when the dielectric layer is identical in the two sides of the film [38]. The hole dimensions are in the intermediate range of few hundred nanometers (50-700nm), which puts our metasurface in between the nanometer-size regime (much smaller than the wavelength) and photonic crystal (larger than the wavelength) regime.

For a given film thickness, this system has 4 degrees of freedom (Figure 2): hole dimensions (xh,yh) and array periodicities (px,py). These 4 parameters determine the resonant frequencies supported by the system - Localized Surface Plasmons, Propagating Surface Plasmons, cavity modes (because of non-negligible thickness), and thus the entire optical response. Naturally, a much wider parameter space is possible where the individual cavity shape is also changed, but this is outside the scope of the current work. Beyond these four parameters, it is possible, of course, to change the film thickness, and this parameter will affect the tuning of the cavity mode resonance.

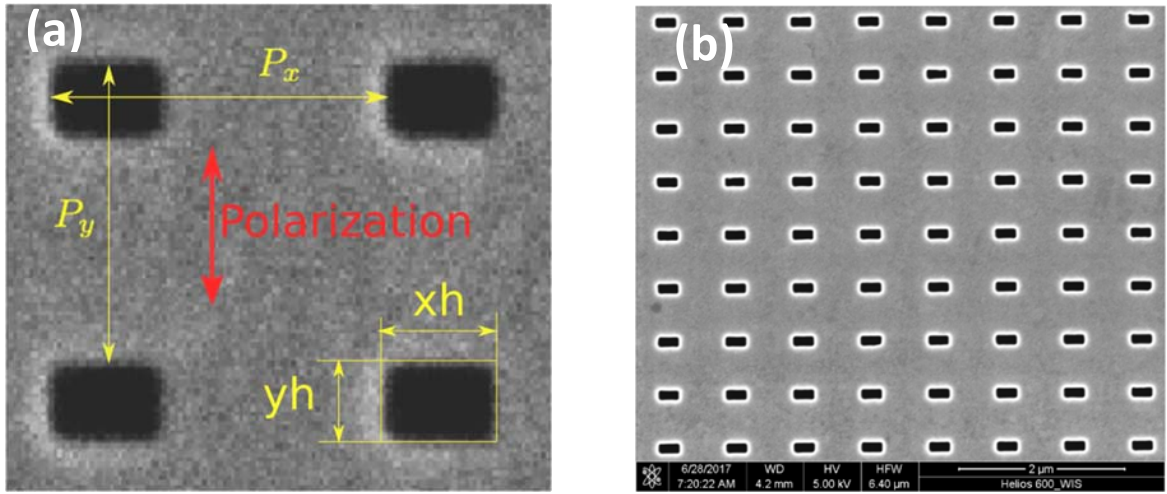

Figure 2 Metasurface array – **(a)** the definition of the parameters used in the optimization, and **(b)** the SEM images of fabricated arrays.

1. **<u>Optimization</u>**

To identify the configuration that gives the strongest FWM signal for arrays of rectangular cavities, we adopted two different strategies: the first is based on a calculation of the linear spectral response of a given configuration, and the second on the direct nonlinear calculation of the FWM signal generated by the array. Clearly, the nonlinear calculation is much more involved, both conceptually and in terms of computing resources and time, so that the comparison of the two may provide interesting insight into the entire optimization process.

   a. Linear response optimization

To find the "best" configuration, we calculated the individual linear transmission spectrum for every configuration by numerically solving the Maxwell's equations by means of 3D Finite Difference Time Domain (FDTD) method using the Lumerical commercial FDTD package [39]. We then used the Genetic Algorithm (GA) optimization approach [40] to find the best one. GA is used extensively in optics, for example pulse shaping for coherent control [41–45], and more recently for plasmonic design [46–50] in the linear regime.

The Genetic Algorithms is inspired by the biological evolution. The process is initialized by randomly selecting an initial population, the "first generation" of "individuals", where each individual represents a point in the parameter space (the values of the parameters are its "genes"). The algorithm proceeds by evaluating each individual according to the specified objective function. Those individuals receiving the larger values (more "fit to survive") are then used to construct the next generation of individuals, by using various "biological" operators: "mutation", when the genes of a single individual are randomly modified within a predefined range; and "cross-breeding", when the genes of two individuals ("parents") are mixed to create a new individual

("offspring"). This creates a continuous process of improvement from one generation to the next, hopefully converging to an optimized individual (convergence is reached when there is no more improvement when moving to the next generation), which is presumably the best one within the defined parameter range. The genetic algorithm is an efficient optimization strategy in cases where the derivative of the function is not known, or does not exist, or when the function is not even known to be continuous as it is being numerically calculated at discrete points.

Motivated by the conventional wisdom stemming from many years of nonlinear optics with molecular samples, we wish to generate resonance enhancement by tuning the samples to the laser frequencies. Thus, the quantity we wish to optimize, namely the target function in the language of optimization procedures, is the product of the transmission coefficients (normalized to the relative area occupied by the holes) at the frequencies participating in the nonlinear optical process.

For our FWM process, where two photons at frequency $\omega_1$ interact with one photon at frequency $\omega_2$ to generate a photon at $\omega_{FWM}$ this target function is $F(x) = T^2(x, \omega_1) T(x, \omega_2) T(x, \omega_{FWM})$, where $T(x, \omega)$ is the transmission at frequency $\omega$ for configuration x. As noted, the input frequencies are fixed, and the 4 component vector $x = (x_h, y_h, P_x, P_y)$ defines a set of parameters for a given geometry. The sample thickness was kept fixed and equal to 320nm. The GA optimization process finds the configuration giving the highest value of the target function $F(x)$.

Following the above procedure, the best configuration, the one with the highest value of the target function, was found to be for x₁=(400,100,860,620), where dimensions are listed in units of nm. The resulting normalized linear transmission spectrum is depicted in Figure 3.

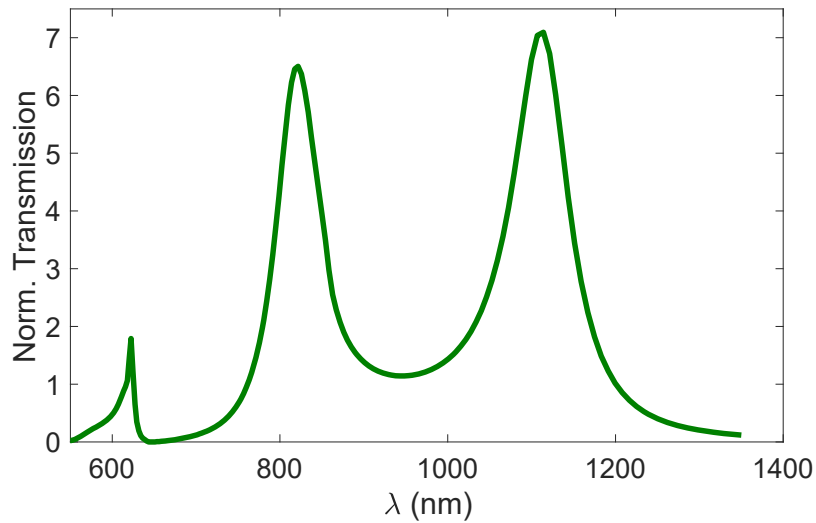

Figure 3 The calculated spectrum of the linearly optimized winning configuration: x₁=(400,100,860,620)

One can immediately realize that the optimization process found a configuration that gives rise to a linear spectrum with pronounced resonances at the required frequencies. Intuitively, and in analogy to the conventional insight gained in molecular spectroscopy, this configuration is expected to yield high NL response, as it is triply resonant. Note that it is not a-priori obvious that a configuration that is triply resonant at the relevant frequencies even exists. In this configuration, the high-frequency mode at 622nm is a radiative SPP-Bloch mode, resulting from the coupling (via the cavities) between the SPPs from the opposite sides of the film. The sharp dip at 646nm is the Wood's anomaly. The peak at 800nm is a cavity mode, also called Fabry-Perot type resonance. These three features create what is known as Fano-type resonant behavior. The lowest frequency resonance is an LSP mode.

b. Direct Nonlinear FWM optimization

As noted, an alternative approach to the optimization is to repeat the GA procedure, but this time to use the directly calculated FWM response as the target function that is to be optimized. As we have shown, this nonlinear response can be calculated within the commercially available FDTD package Lumerical. Naturally, the numerical calculation is more involved, and the optimization is more time and resource demanding.

A-priori, the direct nonlinear optimization is expected to be more accurate than the one based on the naïve approach, but since the linear approach is based on years of experience with molecular systems, our expectations were that both optimization methods would give similar results. To our great surprise, this was not the case.

The best directly optimized configuration was found to be $x_2$=(260,130,790,630). Its calculated linear transmission spectrum is shown in

To substantiate a (solid blue line), together with the spectrum of the linearly optimized winning configuration $x_1$=(400,100,860,620).

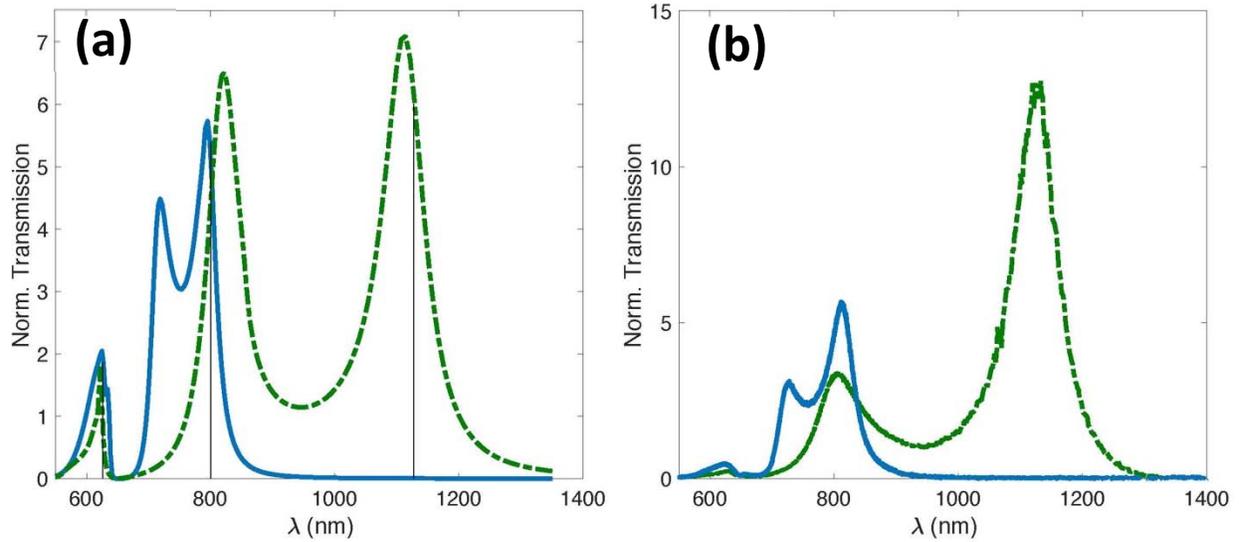

Figure 4 **(a)** The linear transmission spectrum (solid blue line) of the nonlinearly optimized winning configuration $x_2$=(260,130,790,630) with two resonances and without any resonance at the third frequency. For comparison, the transmission spectrum of the linearly optimized winning configuration $x_1$=(400,100,860,620) is shown by the dashed green line. **(b)** The measured linear transmission of the two configurations.

While the parameters of the two winning configurations do not seem to be very different, the linear transmission spectrum is very different: the linearly found configuration $X_1$ gives rise to the expected triply resonant spectrum, whereas the directly nonlinearly optimized configuration $X_2$ yields a spectrum that is only doubly resonant and does not exhibit a peak at the third frequency. Thus, the intuitive expectation will be for a higher FWM signal for configuration $X_1$. **However, the theoretically calculated FWM intensity for $X_2$ is 3 times higher than the one calculated for configuration $x_1$.** When we checked the value of $F(x_2)$ the target function of the linear optimization methodology, it is found to be many orders of magnitude smaller than the value obtained for $x_1$.

To conclude this discussion of our optimization schemes: we have performed two different GA optimization procedures – linear and nonlinear. For both schemes we found and compared the winning configurations – the linear optimization yielded a triply resonant configuration which gave significant enhancement of the FWM signal whereas the direct nonlinear optimization resulted in a configuration that gives rise to a nonintuitive linear transmission spectrum with peaks only at two frequencies. However, to our initial surprise, this configuration gives rise to a stronger FWM signal.

## 2. Fabrication and FWM measurements

We fabricated the two configurations and measured their linear and nonlinear FWM response. As noted above, we used freestanding gold films to eliminate the non-resonant background generated in the much thicker substrate. The linear transmission spectra are depicted in figure 4b – the green line for the linearly optimized configuration $X_1$ , and the blue line for the directly nonlinear optimized configuration $X_2$. We find good agreement with the theoretically calculated spectra. **For both configurations, the generated FWM signal was measured to be similar**. Note, however, that the calculated FWM response for the nonlinearly optimized configuration $X_2$ was higher by a factor of 3, a surprising result in view of the non-intuitive transmission spectrum with only two peaks.

To substantiate the optimality of the obtained configuration, we calculated (and later fabricated and measured) sets of configurations in the neighborhood of the optimal ones, changing the hole dimensions $(x_h, y_h)$. For the configuration $X_1$ the result is shown in Figure 5 and figure 7a. Figure 5a depicts the value of the target function in the linear optimization scheme for a range of parameters around the winning configuration $X_1$. It is not surprising that the value at exactly $X_1$ is the maximum. Figure 5b depicts the calculated FWM response for the configurations with the same range of parameters. One can clearly see that the calculated FWM response closely follows the distribution of the values of the linear target function, strongly suggesting that the linear optimization scheme is a very rational approach to finding the maximal FWM response. Figure 7a shows the values of the measured FWM intensity of 6 configurations ($X_1$ itself and additional 5) as vertical bars, superimposed on the same calculated values of the nonlinear signal as in Figure 5b, demonstrating good agreement.

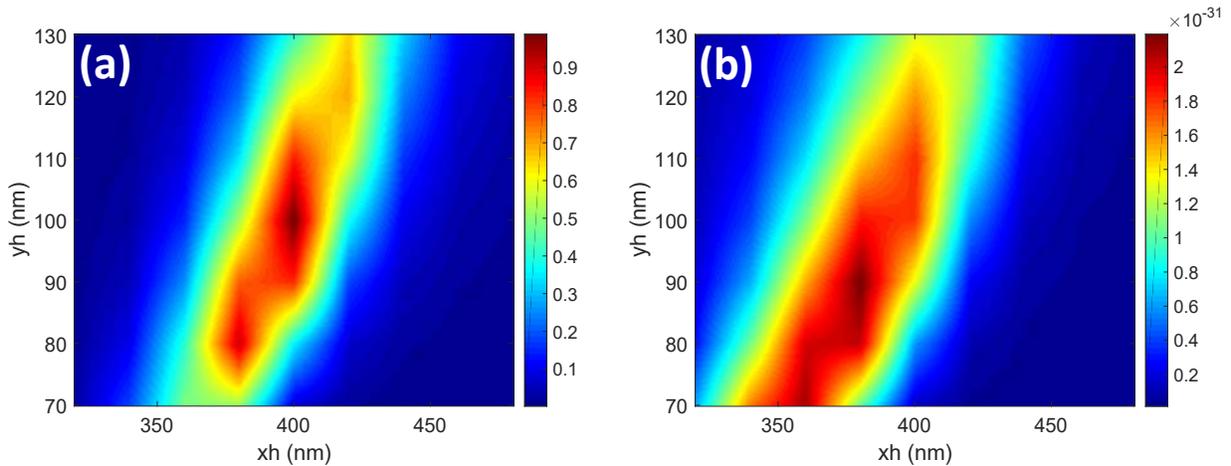

Figure 5 **(a)** The values of the target function, $F(x) = T^2(x, \omega_1) T(x, \omega_2) T(x, \omega_{FWM})$, calculated over a range of parameters around the winning configuration $X_1$. **(b)** Calculated FWM signal over the same range of parameters, indicating strong similarity to the linear prediction.

The same procedure was repeated for the directly optimized result, and a set of configurations around $X_2$ was calculated, as depicted in figure 6. As expected, the GA indeed found the optimal configuration within this range of parameters. Figure 7b shows the values of the measured FWM intensity of 8 configurations (X$_2$ and 7 additional ones), again showing reasonably good agreement with calculation.

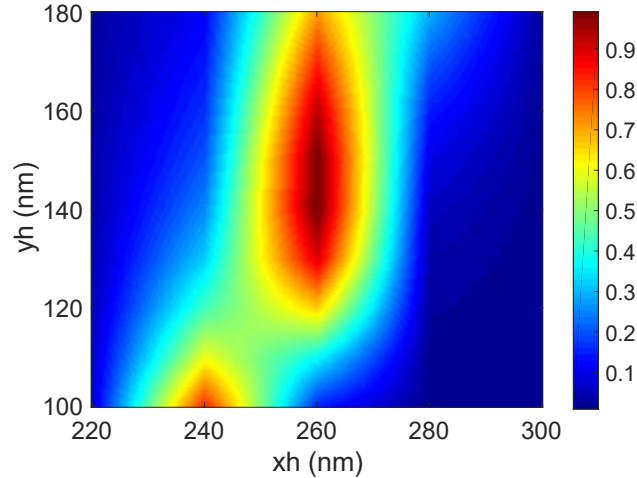

Figure 6  The calculated FWM signal over a range of parameters around the configuration X$_2$.

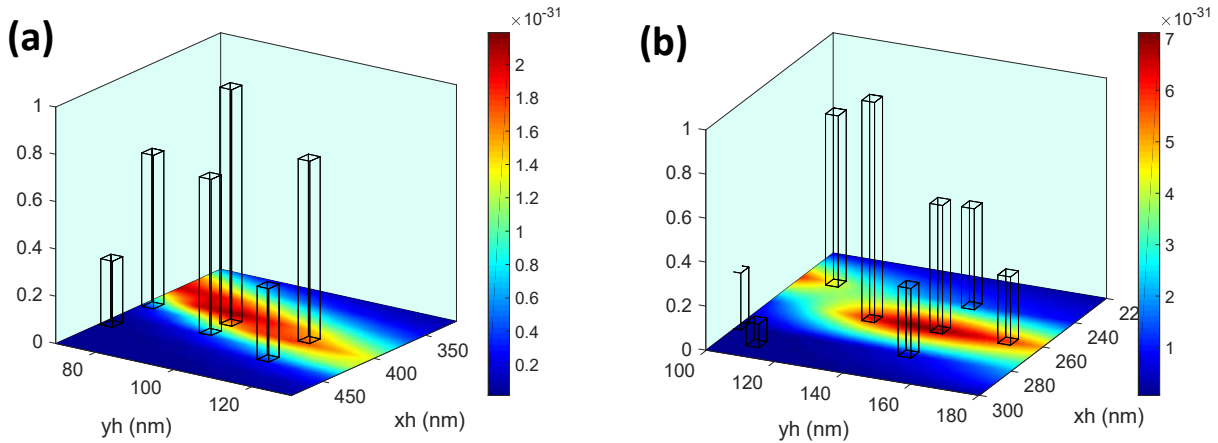

Figure 7 The calculated FWM signal over a range of parameters for **(a)** the linearly optimized and **(b)** the directly optimized configurations. The height of the vertical bars in each figure represents the measured nonlinear signal intensity of the corresponding configurations (normalized to the highest value in each set).

# Discussion

The optimization that is based on the NL calculation improved the resulting FWM enhancement, and this is not surprising, as the calculation takes into account the actual NL FWM process. What is surprising is that the corresponding linear transmission spectrum does not match our expectations for triply resonant behavior. The linear optimization which is based on the far field transmission found a significant enhancement, but missed a configuration which yields high FWM signal but without the triple resonance enhancement. To analyze these results, we must look at other factors influencing the magnitude of the FWM signal. This counterintuitive observation that a seemingly doubly resonant configuration yields a nonlinear signal that is higher than a triply resonant configuration warrants a special discussion.

To start this discussion, we note that the linear transmission spectrum is calculated and observed in the far field, namely not very close to the sample. The nonlinear optical interaction between the three incoming waves via the material nonlinear susceptibility, $\chi_3$, generates the outgoing FWM. This interaction occurs exclusively in the active region at the metal-cavity interfaces and right next to them. This simple fact introduces an important limitation on the use of far field observations since any non-radiating modes that may exist in the near field will not be seen in the far field transmission spectrum. This critical importance of the contribution of dark modes to nonlinear optical interactions had been noted before in the context of Fano resonances and their contribution to NL interactions [51,52]. Thus, the field intensity (or mode distribution) in the near-field is the dominant factor which determines the strength of the FWM interaction. Indeed, when looking at the field distribution in the three constituent frequencies of the FWM process for configurations $X_1$ and $X_2$ (figure 8), the near field at the anti-Stokes frequency (1127 nm) in $X_2$ is completely invisible in the far field spectrum. Therefore, in order to predict the nonlinear enhancement from the linear properties, one must calculate and use field distribution at and near the metasurface itself.

As had already been observed many years ago, when several terms (pathways) contribute to a nonlinear process, their constructive (destructive) interference is important [53]. For molecules, it is the phase resulting from the quantum mechanically interfering wave functions for each transition. For nanostructures it is the Spatial Overlap Integral (SOI) of the field mode distribution inside the sample. This factor is not present in Angstrom size objects such as molecules but is very relevant in structures of dimensions which are comparable to (even if smaller than) the optical wavelength in the material. The nonlinear signal generated inside a metasurface can be expressed [54–56] using the Lorenz reciprocity theorem by

$$NL \approx \iiint_{\text{Gold}} \chi^{(3)} E_1^2 E_2^* E_{FWM}^\dagger dV$$

where $E_1, E_2$ are the input beam local electric fields, and $E_{FWM}^\dagger$ is the time-reversed field at the FWM frequency (Lorenz reciprocity principle). These values vary as function of the spatial coordinates throughout the metasurface. The integration is performed over the volume where the nonlinear susceptibility is nonzero, i.e. over the points in space inside the gold (and in particular excluding the nano holes). The fields $E_1, E_2, E_{FWM}$ are complex-valued, and so there can be cancellations in the integral, because of destructive interference. Note that the calculation of the SOI is a linear calculation where for $E_1$ and $E_2$ we use the actual incoming fields, and for the $E_{FWM}$ we use a field launched into the sample with the other two incoming waves but at the frequency of the generated FWM signal.

Using the expression for the SOI, our calculations show that within the relevant parameter range, the mode overlap for the NL-optimized configuration is similar to that of the linearly optimized configuration. Thus, as can be seen in Table 1, this overlap integral which is a result of the linear calculation, is a good predictor for the configuration that gives the highest FWM signal, in fact, better than the target function F(x) which was used for the optimization (see below).

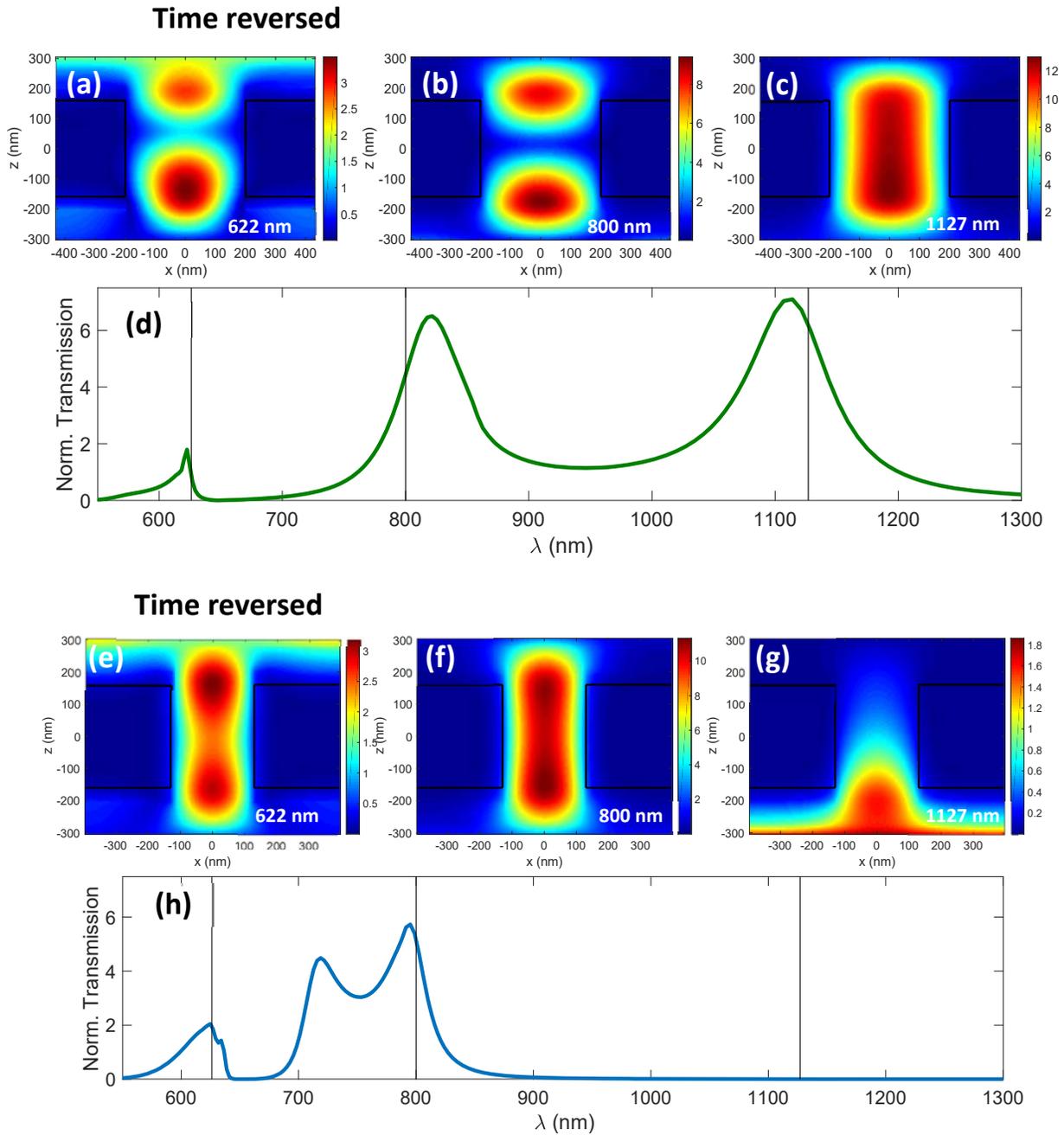

Figure 8 Cross-section at the hole center of the mode distribution for the two configurations at three different frequencies –2 incoming and the FWM. The images are obtained via linear FDTD simulation. **(a)-(d)**: linearly optimized configuration $X_1$; **(e)-(h)** direct FWM optimized configuration $X_2$. The cross-section (XZ plane) shown is perpendicular to the direction of the polarization (Y axis), which is also the shorter side of the hole. The metal boundaries are shown by the black thick lines. The color code (different for each panel) indicates the local field enhancement factor relative to the input field.

|  | Linearly optimized (Configuration $x_1$) | Directly optimized (Configuration $x_2$) |
|---|---|---|
| Calculated NL signal | $2.2 \times 10^{-31}$ | $7.2 \times 10^{-31}$ |
| Spatial Overlap Integral (arb. Units) | $1.9 \times 10^{-14}$ | $6.9 \times 10^{-15}$ |
| $F(x)$, linear prediction (arb. Units) | 160 | 0.02 |

Table 1 The values of the calculated nonlinear enhancement (using 3D FDTD), compared with SOI and the linear prediction function for the two configurations $x_1$ and $x_2$. The strength of the nonlinear signal is equivalent for the two configurations, the linear prediction function gives 4 orders of magnitude advantage to the linear far-field optimization, while the SOI metric gives the same order of magnitude value, which is much closer to the calculated nonlinear signal.

In order to investigate the overlap integral further, we have calculated its value in the neighborhood of the two configurations identified as optimal by the two methodologies, and compared it with the corresponding value of the NL enhancement. The results are shown in figure 9.

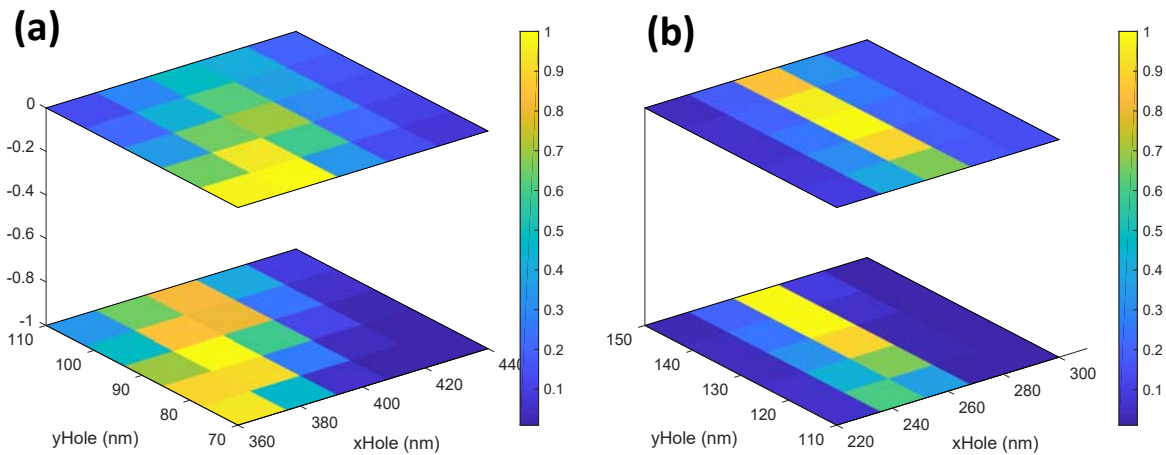

Figure 9. comparison of the SOI to the calculated FWM signal, in both cases the top plane is the SOI and the bottom plane is the calculated FWM signal intensity **(a)** for the neighborhood of $X_1$, the optimal configuration as found by the linear optimization scheme. **(b)** for the neighborhood of $X_2$, the optimal configuration as found by the direct calculation.

Note the good match in Figure 9b, which indicates that the SOI is a reasonable approximation for the full nonlinear calculation. While the SOI does not exactly match the calculated FWM optimized signal, it approximates it very well. Considering the fact that the Nonlinear calculation is much more involved and demanding in terms of time and computational resources, for large parameter spaces one may consider running the GA optimization using a linear calculation with the SOI as the target function.

In conclusion, we have introduced multivariable optimization as a tool for finding configurations that yield an optimized FWM signal. We used two different strategies in our genetic algorithm optimization schemes, where in one case the target function to be optimized is the product of the transmission at the three relevant frequencies, and indeed this linear optimization yielded the anticipated triply resonant configuration that gave rise to significant enhancement of the FWM signal. A different approach was to actually calculate the FWM signal, and use this calculated result as our target function in the optimization procedure. This direct nonlinear optimization yielded a higher FWM signal, but showed only a doubly resonant transmission spectrum – a counterintuitive result. We explain this result in terms of local mode, including the dark mode, excitations within the metasurface, and derive reasonable agreement between predictions and observations. The last point to be considered is hybrid optimization scheme, where local fields are calculated linearly; the resulting spatial overlap integral is then derived and used as the target function in the optimization scheme.

# Methods

### Design optimization

The setup of the genetic algorithm is given in Table 2. We used standard implementation in MATLAB. For FDTD simulations we used Lumerical FDTD commercial software with uniform mesh of size 2.5nm in all axes, and Johnson & Christie (J&C) model for gold. For the nonlinear FDTD simulations, a new nonlinear material was created based on J&C, with $\chi_3 = 10^{-18}$ m$^2$/V$^2$.

| Generation size | Elite | Cross breeding | Mutation | Parameter type | Parameter ranges (nm) | | | | |
|---|---|---|---|---|---|---|---|---|---|
|  |  |  |  |  |  | $x_h$ | $y_h$ | $P_x$ | $P_y$ |
| 20 | 20% | 50% | 30% | Integer, step sizes of 10 nm | min | 80 | 80 | 70+$x_h$ | 70+$y_h$ |
|  |  |  |  |  | max | 700 | 400 | 700+$x_h$ | 700+$y_h$ |

Table 2 Genetic algorithm parameters

### Free-standing Au film preparation

A free-standing 320 nm thickness gold film was created, in which the hole-arrays were later fabricated. First, Silicon Nitride mask was deposited on one side of a [100] silicon wafer by plasma-enhanced chemical vapor deposition (PECVD). Undesired silica created in this process was removed by Buffered oxide etch (BOE). The wafer was divided into rectangles by cleaving it along the crystal lattice using a diamond scribe. In the next step, deposition of 10 nm chromium layer, followed by 325 nm gold layer was performed on the other side of the silicon wafer by e-beam evaporation. The silicon and the chromium layers were then removed under the desired area: using 45% wt. KOH solution at 80 C for the chemical etching of the silicon, and then by dipping the sample for a few seconds into HClO4 37% for removing the chromium. The selectivity of the removal was achieved due to the Silicon Nitride mask deposited in the first stage. The process is schematically shown in the Supplementary Material of [21].

### Hole array fabrication

The milling of the hole-arrays, according to the designs obtained earlier, was carried out using the Focused Ion Beam (FIB) Helios NanoLab 600 machine. The apertures used for the milling: 28 pA .

### Linear spectra measurements

The light source for the linear measurements was a 20W Quartz-Tungsten-Halogen Lamp (Newport, 66310 QTH). The incoming beam was normal to the sample. Two separate spectral ranges were measured using different detectors:

450-1000 nm: CCD (Jobin-Yvon, Symphony) coupled to a spectrometer (Jobin-Yvon, Triax 180);

1000-1450 nm: InGaAs photodiode (Thorlabs, SM05PD5A) coupled to a spectrometer.

The spectra acquired with the different detection systems were each normalized to the light transmitted through an open area equal to the integrated area of the exposed array, and stitched for complete spectral visualization.

### Nonlinear signal measurements

The fundamental beam at $\lambda_1 = 800$nm was provided by a Chirp Pulse Amplifed Ti:Sapphire laser (Spectra Physics Spitfire pumped by a Mai Tai) of duration 60 fs and repetition rate 1 kHz. The combined power of both input beams is 300 µW. The second beam was generated by an Optical Parametric Amplifier (OPA) pumped by the same laser. The FWM signal was measured at $\lambda_{FWM} = 622$nm. The OPA frequency was tuned accordingly to $\lambda_2 = 1127$nm. The optical path of the $\lambda_1 = 800$nm beam was adjusted by a delay line (DL) to temporally overlap with the $\lambda_2$ beam at the sample. The experimental setup is schematically shown in the Supplementary Material of [21].

# Acknowledgement

We thank Ori Avayu and Tal Ellenbogen for their help with the linear transmission measurements. We thank Peter Nordlander for useful discussions and for bringing relevant earlier work to our attention. EA acknowledges the support of the Koshland Foundation. This work was supported, in part, by the Minerva Foundation and by the ICORE program of the Israel Science Foundation.